\documentclass[prl,aps,twocolumn,showpacs,preprintnumbers,amsmath,amssymb]{revtex4}

\usepackage{graphicx}
\usepackage{dcolumn}

\begin{document}

%\preprint{APS/123-QED}
\title{Electron and phonon band-structure calculations
for the antipolar SrPt$_{3}$P antiperovskite superconductor:
Evidence of low-energy two-dimensional phonons
}

\author{Chang-Jong Kang$^1$, Kyo-Hoon Ahn$^2$, Kwan-Woo Lee$^{2,3}$,
B. I. Min$^{1,}$}
\email[mail to : ]{bimin@postech.ac.kr}
\affiliation{$^1$Department of Physics, PCTP,
	Pohang University of Science and Technology, Pohang 790-784, Korea \\
    $^2$ Department of Applied Physics, Graduate School,
    Korea University, Sejong 339-700, Korea \\
    $^3$ Department of Display and Semiconductor Physics,
    Korea University, Sejong 33-700, Korea
}

\date{\today}

\begin{abstract}
SrPt$_{3}$P has recently been reported to exhibit superconductivity
with $T_{c}$ = 8.4 K.
To explore its superconducting mechanism, we have performed
electron and phonon band calculations based on the
density functional theory, and found that the superconductivity
in SrPt$_{3}$P is well described by the strong coupling
phonon-mediated mechanism.
We have demonstrated that superconducting charge carriers
come from $pd\pi$-hybridized bands between Pt and P ions,
which couple to low energy ($\sim 5$ meV) phonon modes confined
on the $ab$ in-plane.
These in-plane phonon modes, which do not break antipolar nature of
SrPt$_{3}$P, enhance both the electron-phonon
coupling constant $\lambda$ and the critical temperature $T_{c}$.
There is no hint of a specific phonon softening feature in the phonon
dispersion,
and the effect of the spin-orbit coupling on the superconductivity
is found to be negligible.
\end{abstract}

\pacs{74.20.Pq, 74.25.Kc}

\maketitle
%==============================================================================

New antiperovskite superconductors have been recently discovered
in a family of Pt-based phosphides $A$Pt$_{3}$P ($A$ = Sr, Ca and La)
with $T_{c}$ = 8.4 K, 6.6 K and 1.5 K, respectively~\cite{Takayama12}.
SrPt$_{3}$P, which has the highest $T_{c}$ among $A$Pt$_{3}$P,
was classified as a strong coupling $s$-wave superconductor
with $2\Delta/k_BT_{c}\approx 5$~\cite{Takayama12}.
This system is reminiscent of other antiperovskite superconductors,
MgNi$_{3}$C with $T_{c} \sim 8$ K~\cite{He01}
and CePt$_{3}$Si with $T_{c}$ = 0.75 K~\cite{Bauer04}.
$A$Pt$_{3}$P has a strongly exchange-enhanced paramagnetic element Pt
in common with MgNi$_{3}$C that has a magnetic element Ni.
The superconductivity in MgNi$_{3}$C was explained well by
the phonon-mediated mechanism~\cite{Shim01}.
On the other hand,  CePt$_{3}$Si is a heavy fermion superconductor
having polar nature due to lack of inversion symmetry.
While $A$Pt$_{3}$P has inversion symmetry,
it also possesses spontaneous polarizations but with antipolar nature.
The superconducting mechanism in CePt$_{3}$Si is still under
intensive investigation~\cite{Saxena04}.

\begin{figure}[b]
\includegraphics[width=6.5 cm]{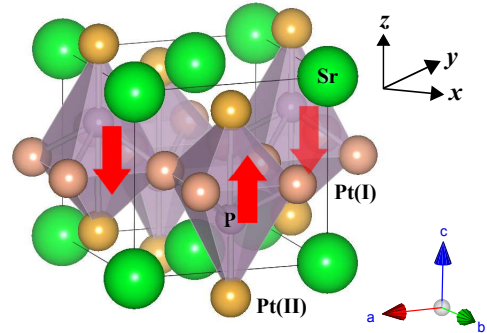}
\caption{(Color Online)
Crystal structure of SrPt$_{3}$P. Pt ions have two symmetry sites,
4e and 2c, at which Pt(I) and Pt(II) are located, respectively.
Due to the displaced P ion in the Pt$_{6}$ octahedron, the
electric polarization emerges as marked by red arrows.
Because the polarizations are antiparallel,
SrPt$_{3}$P has antipolar nature.
Local $x$, $y$, and $z$ coordinates for analyzing
partial characters of bands and DOSs are presented.
}
\label{struct}
\end{figure}

Distinctly from the cubic antiperovskite MgNi$_{3}$C,
SrPt$_{3}$P crystallizes in a tetragonal structure
with a space group ($P$4/$nmm$) (see Fig.~\ref{struct}).
Phosphorous (P) anions are located inside Pt$_{6}$ octahedra.
There are two types of Pt ions.
The Pt(I) sits on 4e symmetry site
in the in-plane of the system, while the other Pt(II) sits
on 2c symmetry site at the out-of-plane position.
In-plane Pt(I) ions form a square lattice
and have equal distances from the central P anion,
which is located closer to one of the Pt(II) ions along the $c$-axis.
Due to the displaced P ion inside the distorted Pt$_{6}$ octahedron,
the spontaneous electric polarization emerges in
SrPt$_{3}$P, as shown in Fig.~\ref{struct}.
The polarizations in SrPt$_{3}$P are ordered in an antiparallel manner
within the unit cell, and so SrPt$_{3}$P has antipolar nature.

%------------------------------------
%------------------------------------

Based on the measurements of specific heat and Hall resistivity,
Takayama \emph{et al.}~\cite{Takayama12} proposed existence
of low-lying phonons and multiple Fermi surfaces in SrPt$_{3}$P.
They suspected that the strong spin-orbit coupling (SOC)
of 5$d$ electrons in Pt might play some roles in creating the multiple
Fermi surface pockets.
Then they suggested that the low-lying phonons couple strongly to
the multiple Fermi surfaces,
which would produce the relatively high $T_{c}$ in SrPt$_{3}$P.
Nekrasov \emph{et al.}~\cite{Nekrasov12}, thereafter, calculated
the band structure of SrPt$_{3}$P using the linearized muffin-tin orbital
(LMTO) method and
confirmed the multiple Fermi surfaces in this system.
They found that the multiple Fermi surfaces have a quite complicated
three-dimensional (3D) topology,
indicating an intricate superconducting gap structure.
Therefore, in order to clarify the superconducting mechanism in SrPt$_{3}$P,
it is urgently demanded to examine phonon dispersions and the
electron-phonon coupling strength in SrPt$_{3}$P.

%-----------------------------------------------
\begin{figure}[t]
\includegraphics[width=8.5 cm]{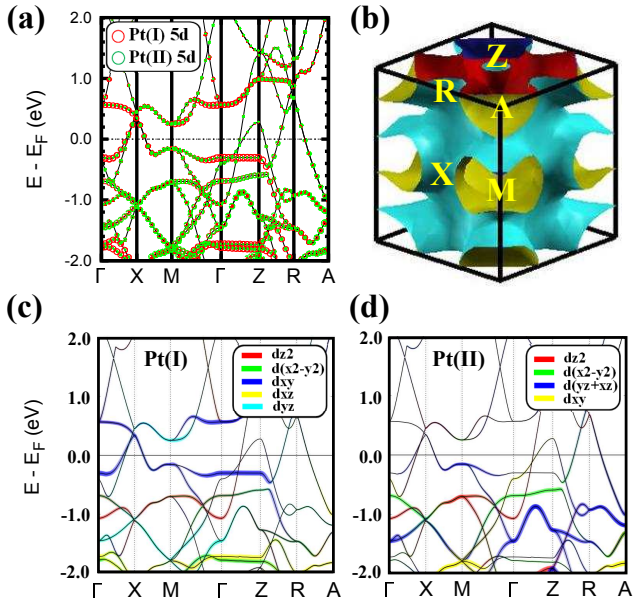}
\caption{(Color Online)
(a) Band structures and (b) Fermi surface of SrPt$_{3}$P
(WIEN2k results without considering the SOC).
The band dispersions near $E_F$ and Fermi surfaces are essentially
the same between the SOC and without the SOC scheme,
indicating a negligible effect of the SOC.
(c) Projected Pt(I) and (d) Pt(II) band character plots.
Pt(I) $d_{xy}$ and Pt(II) $d_{xz+yz}$ states crossing $E_F$
are important in the superconductivity of SrPt$_{3}$P.
}
\label{bandfs}
\end{figure}
%-----------------------------------------------

%------------------------------------
\begin{figure}[t]
\includegraphics[width=8.5 cm]{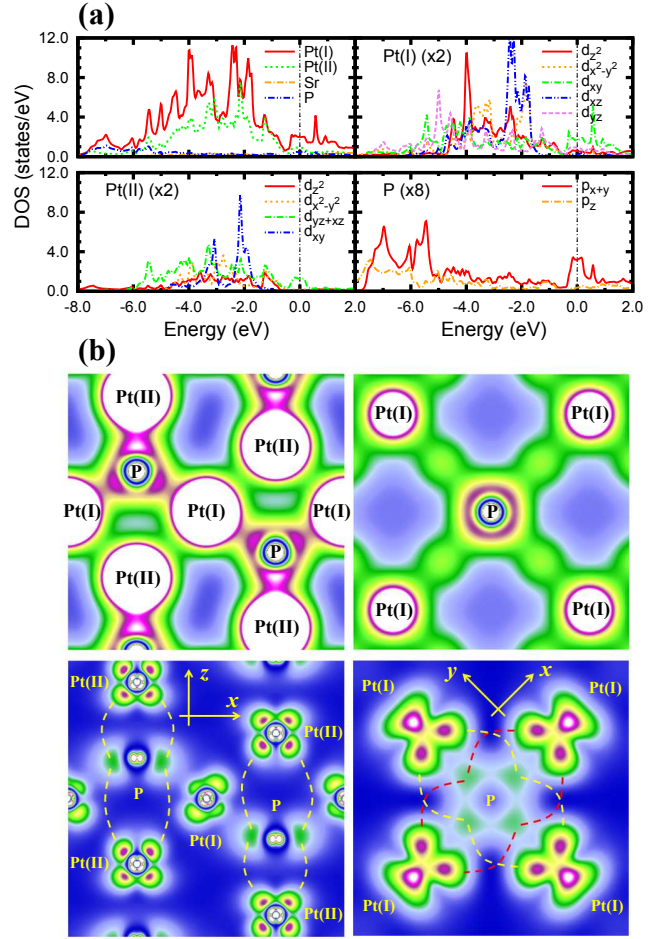}
\caption{(Color Online)
Density of states (DOS) in the no-SOC scheme of WIEN2k.
(a) Projected and partial DOSs per unit cell for Sr, Pt(I), Pt(II), and P.
(b) (Top) Total charge densities on the $xz$ and $xy$ planes.
The dominant bondings between the Pt and P ions are $pd\sigma$-states,
which are located far below $E_F$.
(Bottom) Charge densities near $E_F$ (from $-0.2$ eV to $E_{F}(=0)$)
on the $xz$ and $xy$ planes,
which manifest $pd\pi$-antibondings between Pt and P ions.
The dotted yellow and red lines are guides for $\pi$-antibonding.
}
\label{dosden}
\end{figure}
%------------------------------------

In this letter, we have investigated electronic structures and
phonon dispersions of SrPt$_{3}$P to explore
its superconducting mechanism.
We have found that the phonon-mediated mechanism
describes well the superconductivity in SrPt$_{3}$P,
and superconducting carriers arise mostly from
the in-plane $pd\pi$-hybridized states between Pt and P ions,
which are strongly coupled to specific low energy phonon modes
confined on the $ab$ in-plane.
Those two-dimensional (2D)-like vibrations do not break
antipolar nature in SrPt$_{3}$P.
In addition, we have discussed
why SrPt$_{3}$P has the highest $T_{c}$ among the family
of Pt-based superconductor $A$Pt$_{3}$P ($A$ = Sr, Ca, and La).

For the electronic structures, we have used both
the all-electron (WIEN2k)~\cite{Wien2k} and
the pseudopotential (Quantum Espresso)~\cite{Q-espresso} band methods
in the generalized gradient approximation (GGA),
while, for phonon dispersion calculations, we have used
the linear response method~\cite{Baroni01} implemented in the latter
code~\cite{Ultra}.
Since the Wilson ration for SrPt$_{3}$P is close to one~\cite{Takayama12},
correlation effects will not be large, and so the standard
density functional theory calculation will be sufficient
to describe the system.
Structural parameters of SrPt$_{3}$P are employed
from the experiment~\cite{relax}.

Figure~\ref{bandfs} shows the band structures and the Fermi surface of
SrPt$_{3}$P without the SOC.
We have confirmed that the band structures with and without the SOC are
essentially the same,
especially near the Fermi level ($E_F$), and so Fermi surfaces also
become almost the same.
This feature is contrary to the expectation
that the large SOC in Pt would play an important role
in the superconducting properties of SrPt$_{3}$P~\cite{Takayama12}.
There are two bands crossing $E_F$. The band that forms a hole pocket
Fermi surface (FS) around X comes mainly from in-plane Pt(I) $d_{xy}$ states,
as seen in Fig.~\ref{bandfs}(c).
Two nearly flat bands above and below $E_F$ along M$-\Gamma-$Z also
correspond to these states. On the other hand, the dispersive band
that makes a hole FS around R
has more Pt(II) $d_{xz+yz}$ characters,
which arise from antibonding states between Pt(II) and P.
A very dispersive band crossing $E_F$ along $\Gamma-$Z arises
from the strong antibonding states between Pt(II) $5d$ and P $3p$
along the c-axis.
This free electron-like feature along the $c$-axis is supposed to trigger
2D-like in-plane phonon vibrational modes.

Figure ~\ref{dosden}(a) shows the partial density of states (DOS)
of SrPt$_{3}$P.
The DOS at $E_F$ has contributions largely from
Pt(I) $d_{xy}$, Pt(II) $d_{yz+xz}$, and P $p_{x,y}$ states.
The pair of Pt(II) $d_{z^2}$ and P $p_{z}$ orbitals
and that of Pt(I) $d_{x^2-y^2}$ and P $p_{x,y}$ orbitals form
$pd\sigma$-bondings, as shown in the top of Fig.~\ref{dosden}(b),
respectively,
and so those states are located far below $E_F$.
Superconducting carriers near $E_F$ have $pd\pi$-antibonding
characters, which consist of pairs of
Pt(I) $d_{xy}$ and P $p_{x,y}$ orbitals
and Pt(II) $d_{yz+xz}$ and P $p_{x,y}$ orbitals
(see Fig.~\ref{dosden}(b) bottom).
Hence, the Fermi surface in Fig.~\ref{bandfs}(b) is made of those
charge carriers in the covalent $pd\pi$-antibonding states of Pt(I)-P
and Pt(II)-P ions.
These two different $pd\pi$-antibonding states
support the proposed multiple band nature of superconductivity
in SrPt$_{3}$P~\cite{Takayama12}.

%------------------------------------
\begin{figure}[t]
\includegraphics[width=8.5 cm]{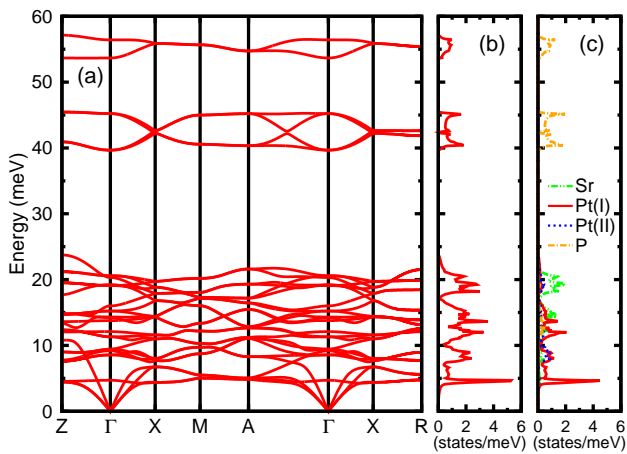}
\caption{(Color Online)
(a) Phonon dispersion, (b) total, and (c) partial phonon DOSs of SrPt$_{3}$P.
Phonon DOS has the peak structure at low energy of $\sim 5$ meV,
which corresponds to the Pt(I) vibration on the $ab$ in-plane.
This mode contributes to the superconductivity dominantly.
}
\label{ph}
\end{figure}
%------------------------------------

%------------------------------------
\begin{figure}[t]
\includegraphics[width=8.5 cm]{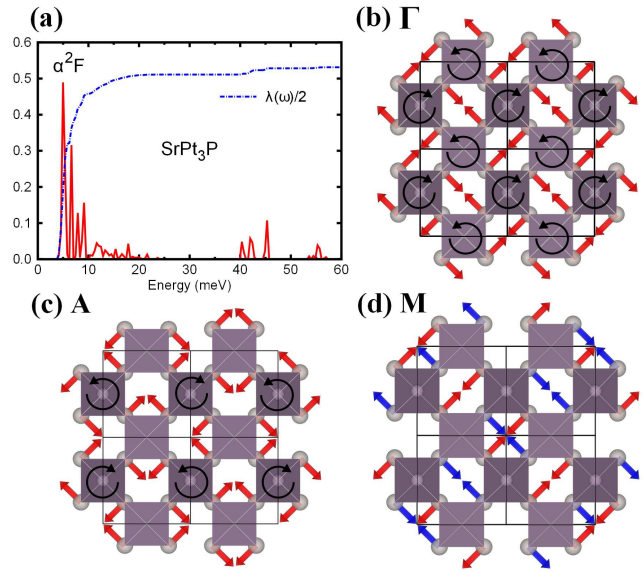}
\caption{(Color Online)
(a) Eliashberg function $\alpha^2F(\omega)$
and the electron-phonon coupling constant $\lambda$ for SrPt$_{3}$P.
$\alpha^2F(\omega)$ has the highest peak at around 5 meV,
which corresponds to the peak position of DOS in Fig.~\ref{ph}(b).
$\lambda$ grows abruptly at the peak position of $\alpha^2F(\omega)$.
(b) Optical phonon normal mode near 5 meV at $\Gamma$,
(c) and that at A, which are viewed along the $c$-axis.
The normal mode at $\Gamma$ corresponds to
rotational modes of Pt$_{6}$ octahedra about the $c$-axis,
while that at A is a variation of that at $\Gamma$.
(d) Acoustic phonon normal mode near 5 meV at M.
Blue and red arrows at M represent
two different acoustic modes, which are degenerate.
Phonon normal modes at M and A yield large $\lambda_{q\nu}$.
}
\label{eliash}
\end{figure}
%------------------------------------

Figure~\ref{ph} shows the phonon dispersion and corresponding
total and partial phonon DOSs of SrPt$_{3}$P~\cite{phonon}.
High energy phonons above 40 meV come solely from the vibration
of P ions, as shown in Fig.~\ref{ph}(c).
The low energy phonons originate from the
mixture of vibrations of the whole ions in SrPt$_{3}$P.
Most interesting is the existence of almost flat optical modes near 5 meV,
which are located even lower than some acoustic modes.
These low energy phonon modes are indeed consistent with the prediction
by Takayama \emph{et al.}~\cite{Takayama12}.
Pt(I) ions play a dominant role in producing
these low phonon energy states, which lead to the highest peak
in the phonon DOS (see Fig.~\ref{ph}(c)).
Note that there is no hint of the specific phonon softening in Fig.~\ref{ph},
such as the imaginary frequency or dip feature, again
unlike the expectation of Takayama \emph{et al.}~\cite{Takayama12}.
Generally, the highest phonon DOS peak in the low energy regime yields
the large electron-phonon coupling constant $\lambda$ and the
critical temperature $T_{c}$.
For comparison, we have also checked the phonon DOSs
of LaPt$_{3}$P and CaPt$_{3}$P, and found that the low energy peak
obtained for SrPt$_{3}$P is much suppressed for
LaPt$_{3}$P and CaPt$_{3}$P.
This explains why SrPt$_{3}$P has the highest $T_{c}$
among $A$Pt$_{3}$P ($A$=Sr, Ca and La).
The reason why SrPt$_{3}$P has the highest peak at low energy phonon DOS
seems to be related to the structural geometry.
Note that crystal ionic radii of Sr$^{2+}$, Ca$^{2+}$, and La$^{3+}$
are 132, 114, and 117.2 pm, respectively~\cite{Shannon76}.
Since the crystal ionic radii of Sr is the largest,
the vibrational mode of Pt(I) ions become mostly localized in SrPt$_{3}$P.
This feature is similar to rattling modes observed
in KOs$_{2}$O$_{6}$~\cite{Hiroi07}.

Figure~\ref{eliash}(a) shows the Eliashberg function $\alpha^2F(\omega)$
and the electron-phonon coupling constant $\lambda$ of SrPt$_{3}$P.
It is seen that $\alpha^2F(\omega)$ has the highest peak at $\sim 5$ meV,
which corresponds to the highest peak position in the phonon DOS.
Accordingly, $\lambda$ grows abruptly at $\sim 5$ meV.
Due to the strong $pd\sigma$-bonding
between Pt(II) and P ions (Fig.~\ref{dosden}(b) top),
the vibrational mode along the c-axis would have high energy,
and so it does not give large $\lambda$.
In contrast, the $ab$ in-plane vibrational mode of Pt(I) is
easily activated, which in turn perturbs the covalent $pd\pi$-antibonding
states between Pt and P ions.
As a consequence, charge carriers in the $pd\pi$-antibonding states
couple strongly with the low-energy phonons so as to produce large $\lambda$.
The in-plane vibration of Pt(I) ions is, therefore,
essential in the superconductivity of SrPt$_{3}$P.
Note that the in-plane vibration of Pt(I) ions does not break
antipolar nature of SrPt$_{3}$P.
Since the in-plane phonon modes of Pt(I) ions couple more strongly
with the $pd\pi$-antibonding state of Pt(I)-P ions than that of Pt(II)-P ions,
a two-gap structure might be possible in SrPt$_{3}$P,
as observed in MgB$_{2}$~\cite{Choi2002}.

The normal modes of almost flat low energy phonons in Fig.~\ref{ph}(a)
are plotted in Fig.~\ref{eliash}(b), (c), and (d),
which correspond to optical modes at $\Gamma$ and A,
and the acoustic mode at M, respectively.
The normal mode at $\Gamma$ in Fig.~\ref{eliash}(b) corresponds to
rotational modes of all the Pt$_{6}$ octahedra about the $c$-axis.
The normal mode at A is considered to be a variation of
that at $\Gamma$. Namely, the normal mode at A also corresponds to
rotational modes of Pt$_{6}$ octahedra, but the neighboring
Pt$_{6}$ octahedron rotates in the opposite direction.
The normal mode at M corresponds to a linear superposition of two degenerate
acoustic modes (red and blue arrows in Fig.~\ref{eliash}(d)).
Four Pt(I) ions get close to or apart from interstitial centers,
whereby electrons are attracted or repelled to be effective in forming
Cooper pairs.  Note that all these low energy phonon
modes do not break antipolar nature, and
strongly couple to $pd\pi$-antibonding states near $E_F$ so as to
yield large $\lambda$.
For example, the acoustic and optical phonon modes at M and A
near 5 meV  yield $\lambda_{q\nu}$'s
of 0.34$-$0.44 and 0.34$-$0.35, respectively.

%------------------------------------
\begin{table}[b]
\caption{Superconducting parameters of SrPt$_{3}$P
($N(E_F)$: DOS at $E_{F}$ per unit cell).
Two values of $T_{c}$ are for two different values
of $\mu^{*}$=0.10 and 0.13,
compared with experimental values of $T_{c}$.
}
\begin{tabular}{c c c c c c c c c}
\hline \hline
& $N(E_F)$ &
& $\lambda$ & $\omega_{\texttt{log}}$ &
\multicolumn{2}{c}{$T_{c}$} & $T_{c}^{\text{exp}}$\\
& (states/eV) & &  & (meV) &
\multicolumn{2}{c}{(K)} & (K)\\
\hline
SrPt$_{3}$P & 3.95 &  &
1.06 & 7.01 & 6.20, & 5.39 & 8.4\\
CaPt$_{3}$P & 4.08 &  &
0.82 & 8.33 & 4.79, & 3.91 & 6.6\\
LaPt$_{3}$P & 4.60 &  &
0.78 & 9.00 & 4.69, & 3.76 & 1.5\\
\hline \hline
\end{tabular}
\label{Tc-table}
\end{table}
%------------------------------------

We now consider the superconducting properties
based on the Eliashberg strong coupling theory~\cite{Eliashberg60}.
The critical temperature $T_{c}$ is obtained
with the McMillan-Allen-Dynes formula \cite{Allen75} :
\begin{equation}
T_{c} = \frac{\omega_{\texttt{log}}}{1.20}
\texttt{exp}\bigg(-\frac{1.04(1+\lambda)}
{\lambda-\mu^{*}-0.62\lambda\mu^{*}}\bigg),
\end{equation}
where $\omega_{\texttt{log}}~\Big(\equiv \texttt{exp}\big(\frac{2}{\lambda}
\int_{0}^{\infty}\frac{d\omega}{\omega}\alpha^{2}
F(\omega)\texttt{ln}\omega\big)\Big)$ is the logarithmic average frequency
and $\mu^{*}$ is the effective Coulomb repulsion parameter.
The estimated superconducting parameters are listed in Table~\ref{Tc-table}.
Total $\lambda$ is obtained to be 1.06,
which is large enough to classify SrPt$_{3}$P as
a strong $s$-wave superconductor.
The Debye temperature is obtained to be 169 K which is consistent
with the measured experimental data 190 K~\cite{Takayama12}.
Then the corresponding $T_{c}$'s of SrPt$_{3}$P are obtained to be
6.20 K for $\mu^{*} = 0.10$ and 5.39 K for $\mu^{*} = 0.13$.
Thus the estimated  $T_{c}$ is in good agreement with the observed $
T_{c}$ of 8.4 K, which indicates that the superconductivity in
SrPt$_{3}$P is described well by the phonon-mediated mechanism.
As shown in Table~\ref{Tc-table},
$T_{c}$'s of CaPt$_{3}$P and LaPt$_{3}$P are lower than $T_{c}$ of SrPt$_{3}$P,
which is mainly due to the reduced $\lambda$'s, as mentioned above
in  Fig.~\ref{ph}.

In conclusion, we have demonstrated that the superconductivity in
SrPt$_{3}$P is described well by the strong coupling
phonon-mediated mechanism, as in MgNi$_{3}$C and MgB$_{2}$.
The superconducting charge carriers correspond mainly to
$pd\pi$-antibonding states of Pt(I)-P and Pt(II)-P ions,
which couple to the low-lying phonon modes
confined on the $ab$ in-plane.
The electron-phonon coupling for the former is stronger than that
for the latter, and so two-gap superconductivity is expected to occur.
The 2D-like phonon modes do not break antipolar nature of SrPt$_{3}$P
and increase the electron-phonon coupling and the critical temperature $T_{c}$.
No evidence of the specific phonon softening is detected,
and the effect of the SOC on the superconductivity
is found to be negligible.

%begin{acknowledgments}
Helpful discussions with Kyoo Kim and Beom Hyun Kim are greatly appreciated.
This work was supported by the NRF (No.2009-0079947)
and the KISTI supercomputing center (No. KSC-2012-C3-09).
The work in KU was supported by the NRF (No. 2012-0002245).
%\end{acknowledgments}


\begin{thebibliography}{99}
\bibitem{Ultra}
    PW91 ultrasoft pseudopotentials were used
	in the Quantum Espresso band and phonon
	calculations, and Gaussian smearing of 0.03 Ry was used
	for the electronic integration.
	We used 10 $\times$ 10 $\times$ 10 k-grid and
	2 $\times$ 2 $\times$ 2 q-grid for phonon calculations, and
	20 $\times$ 20 $\times$ 20 k-grid for the electron-phonon coupling
	calculations.
\bibitem{relax} We have checked that the band structure near $E_F$ and
the Fermi surface do not change much with relaxation of
internal parameters of Pt(II) and P ions.
\bibitem{phonon}
    Since the SOC effect is not dominant in SrPt$_{3}$P,
    we carried out phonon and superconducting calculations
	within the no-SOC scheme.
\end{thebibliography}
\end{document}